\documentclass[11pt, a4paper, onecolumn, twoside]{article}
\usepackage{graphicx}
\def \ket #1{\vert #1 \rangle}

\def \bra #1{\langle #1 \vert}

\begin{document}
% FIRST PAGE
\thispagestyle{empty}
\title{ Ab initio determination of an extended\\ Heisenberg
Hamiltonian in CuO$_2$ layers }
\author{Carmen\ J. Calzado$^1$ and Jean-Paul\ Malrieu 
\\Laboratoire de Physique Quantique. IRSAMC.\\
Universit\'e Paul Sabatier, 31062 Toulouse, France.}
\date{}
\maketitle
\begin{abstract}
Accurate ab initio calculations on embedded Cu$_4$O$_{12}$  
square clusters, fragments of the La$_2$CuO$_4$ lattice, confirm
a value of the nearest neighbor antiferromagnetic coupling
($J$=124 meV)
previously obtained from ab initio calculations on bicentric clusters
and in good agreement with experiment. These calculations predict 
non negligible antiferromagnetic second-neighbor interaction ($J'$=6.5 meV) 
and four-spin cyclic exchange ($K$=14 meV), which may affect the thermodynamic
and spectroscopic properties of these materials.
The dependence of the magnetic coupling on local lattice distortions 
has also been investigated. Among them 
the best candidate to induce a spin-phonon effect seems to be the movement of
the Cu atoms, changing the Cu-Cu distance, for which the variation of the
nearest neighbor magnetic coupling with the Cu-O distance is  
${\Delta J}/{\Delta d_{Cu-O}}\sim$1700 cm$^{-1}$\AA$^{-1}$.
\end{abstract}
\footnote{ On leave from: Departamento de Qu\'{\i}mica F\'{\i}sica. 
Universidad de Sevilla. E-41012. Sevilla. Spain.}
\newpage
\section{Introduction}
Fifteen years after the discovery of the high-T$_c$ superconductivity 
in cuprates, numerous theoretical and experimental studies pay still 
attention to these materials and their parent undoped compounds 
in an attempt to explain their electronic properties\cite{Orenstein}.
Regarding the undoped materials, the $CuO_2$ layers, where  superconductivity
takes place after doping,  can be seen as two-dimensional  spin lattices,
where each Cu atom bears an unpaired electron, which is antiferromagnetically
coupled with the nearest-neighbors(NN). 
The value of this NN coupling has been estimated from Raman
scattering  128$\pm$6 meV\cite{Sulewski,Singh} 
and Neutron diffraction experiments  134$\pm$5 meV\cite{Aeppli, Endoh, Hayden},
assuming a simple Heisenberg Hamiltonian, 
where only NN interactions are considered:
\begin{equation}
H=J\sum_{<ij>,NN} S_i S_j \:\: (J>0)
\end{equation}
where $<ij>$ represents a pair of NN sites. 
However, this simple model does not satisfactorily reproduce
the whole Raman spectra of undoped cuprates 
\cite{Parkinson,Canali,Roger:89,Gagliano, Dagotto, Nori:92},
and extended-Heisenberg Hamiltonians
have been proposed \cite{Eroles:99,Lorenzana:99, Honda:93,Sakai:99}.
The sophistications introduce some of the following effects:\\
- spin-phonon interactions, \\
- next-nearest neighbor (NNN) magnetic couplings, $J '$, and\\
- four-spin cyclic (4SC) exchange, $K$.\\

The spin-phonon coupling, i.e.  the dependence of the magnetic coupling on 
the vibrational distortions of the lattice, has been recently invoked as
possibly responsible for the asymmetry of the $B_{1g}$ peak on the Raman
spectra of cuprates \cite{Eroles:99, Nori:95, Manning, Saenger,Sandvik}. 
The spin-phonon interaction modifies the magnetic coupling $J$ through the
dependence of the hopping integral ($t_{pd}$) and the charge transfer energy
($\Delta_{CT}$) on the Cu-O distance \cite{Calzado:99a, Calzado:99b, Ohta}.
A maximum contribution of $\pm$ 54 meV to the NN magnetic coupling coming
from spin-phonon interactions has been suggested \cite{Eroles:99}, based on 
the linear dependence of $J$ on the Cu-O distance observed in the $M_2CuO_4$
family \cite{Ohta} and the spin-wave approximation.
However, the calculations explicitly including the phonon-spin interaction, 
using an adiabatic approximation for the phonons, \cite{Nori:95}, requiere
unrealistic values of disorder to reproduce the width and asymmetry of 
the $B_{1g}$ peak. 
It seems necessary to introduce additional terms (as NNN coupling and
4SC exchange) to reproduce the structure of the Raman spectra
\cite{Eroles:99}.

The existence of the NNN magnetic coupling $J'$ and the  4SC exchange $K$ 
can be established from a one-band Hubbard model \cite{Maynau:82a,
Maynau:82b, MacDonald:88, MacDonald:90}.
The NNN interactions may be either a second order effect in form of
$\sim t'^{2}/U$, where $t'$ is a second-neighbor hopping integral and $U$ is
the classical on-site Coulomb repusion, or fourth-order effects
scaling as $\sim t^4/U^3$, where $t$ is the NN hopping integral. 
The 4SC exchange is a fourth-order term in the Hubbard model, involving
circulation of the electrons around the square and scaling as $\lambda
t^4/U^3$, where $\lambda$ is a large numerical factor ( $\lambda=$40
\cite{Maynau:82a, Maynau:82b} or $\lambda=$80 \cite{MacDonald:88, MacDonald:90},
depending on the formal writting of the operator), as
shown in early works in quantum chemistry \cite{Maynau:82a, Maynau:82b} and
solid state physics \cite{MacDonald:88, MacDonald:90}.
Recent experiments have shown that four-spin cyclic exchange exists in the 
two dimensional solid $^3$He \cite{Ishida, Roger:98, Misguich}, in the 2D
Wigner solid of electrons formed in a Si inversion layer \cite{Okamoto}
and  in the $bcc$ $^3$He \cite{Osheroff, Roger:83, Cross}.

As was previously shown, oxygen atoms play a crucial role in the spin exchange
between Cu atoms in these materials \cite{Calzado:99a, Calzado:99b,
Martin, Zaanen}. In this context, the one-band model is not suficient to bear 
all the physics of such materials and cannot fix the ratios $J'/J$ and $K/J$.
The multiple parameters contained in an extended Heisenberg Hamiltonian and 
the spin-phonon coupling cannot be univocally fixed from the collective
properties of the material and, as far as possible, a prejudiceless
evaluation of them will be welcome.

In the recent past, ab initio quantum chemical calculations, using large
basis sets and accurate  treatment of the electronic  correlation by means
of extensive configuration interaction (CI) expansions
of the wave functions, have been performed on bicentric clusters
($Cu_2O_7$ and $Cu_2O_{11}$) \cite{Calzado:99a, Calzado:99b},
properly embedded in the Madelung field of the infinite crystal,
crucial to correctly  represent the electronic structure of these
systems\cite{Ohta}. These calculations provided satisfactoy values of $J$ 
(138 meV) and of the first-neighbor hopping integral for the hole-doped system
($t$=-0.55-0.57 eV). In both cases, the evaluation of the effective interaction
goes through the calculation of the spectrum of the dimer.

An extension of this strategy is proposed here which provides an evaluation 
of $J'$ and $K$ from the calculation of the spectrum of four-Cu sites square
embedded clusters. By the way, the transferability of the $J$ value from the
two-center to the four-center clusters will be verified.
To estimate the spin-phonon coupling, the bimetallic cluster $Cu_2O_7$ has
been used, calculating the dependence of the singlet-triplet separation
on different geometry distortions.  

\section{Next-nearest neighbor coupling and four-spin cyclic exchange} 
\subsection{Strategy to extract the effective interactions} 
A square cluster containing four Cu atoms and their nearest twelve in-plane
oxygen atoms (a plaquette) will be used to extract these parameters
(Figure 1). Each Cu atom contains an unpaired electron in an in-plane
$dx^2$-$y^2$-type orbital. For such frame, the four center-four spin model
space is spanned by six neutral determinants.
If one calls $a,b,c$ and $d$ the magnetic orbitals, centered in each Cu atom, 
there are two kinds of determinants with S$_z$=0, the fully spin-alternant
determinants $|a\bar{b}c\bar{d}|$ and $|\bar{a}b\bar{c}d|$ and four
partially-frustrated determinants $|ab\bar{c}\bar{d}|$, $|\bar{a}\bar{b}cd|$,
$|a\bar{b}\bar{c}d|$ and $|\bar{a}bc\bar{d}|$.
The effective Hamiltonian spanned by such a model space can, in full 
generality, be written as:

\begin{tabular}{cccccc}
\cline{1-6}
$|a\bar{b}c\bar{d}|$ & $|\bar{a}b\bar{c}d|$ & $|ab\bar{c}\bar{d}|$ &
$|\bar{a}\bar{b}cd|$ & $|a\bar{b}\bar{c}d|$ & $|\bar{a}bc\bar{d}|$ \\
\cline{1-6}
$-4h-g_4$&$g_4$     &$h$           & $h$           &$h$           &$h$ \\
         &$-4h-g_4$ &$h$           & $h$           &$h$           &$h$ \\
         &          &$-2h-2h'-g'_4$& $g'_4$        &$h'$          &$h'$ \\
         &          &              & $-2h-2h'-g'_4$&$h'$          &$h'$  \\
         &          &              &               &$-2h-2h'-g'_4$&$g'_4$ \\
         &          &              &               &        &$-2h-2h'-g'_4$ \\
\cline{1-6}
\end{tabular}
\\
\\
\\
where the zero of energy is that of the ferromagnetic quintet state 
and the equivalences between different elements are due to symmetric reasons, 
imposed by the structure of the plaquette. For instance, 
the elements $\bra {a\bar{b}c\bar{d}}H^{eff}\ket{ab\bar{c}\bar{d}}$ represents 
the exchange of the spins in $b$ and $c$. In the plaquette, this interaction
must be equivalent to the exchange between $a$ and $d$,
that is, the element
$\bra {a\bar{b}c\bar{d}}H^{eff}\ket{\bar{a}\bar{b}cd}$, and different 
from the $\bra {\bar{a}\bar{b}cd}H^{eff}\ket{a\bar{b}\bar{c}d}$
element, which exchanges the spins on the diagonals.

The six eigenstates of this matrix belong to different
spin-space symmetry irreducible representations. The spectrum can be easily 
written from the basic parameters as shown in Figure 2. There are only four
energy-differences, and then the four parameters can be univocally defined.
If we now perform an accurate calculation of the six lowest eigenstates of
this system, employing the best ab initio techniques, we will obtain four
level spacings which enable us to determine the four desired
effective interactions.
                           
\subsection{Ab initio calculations}
As was previously mentioned, to estimate the NNN and 4SC interactions a 
square cluster containing four Cu atoms and the first 
twelve in-plane oxygen atoms has been considered, where all the atoms are
treated explicitly. The  most internal electrons of the Cu atoms 
($1s^22s^2p^63s^2$) have been replaced by an
effective core potential and the rest of the electrons
($3p^6d^9$ for Cu$^{+2}$ and $1s^22s^2p^6$ for O$^{-2}$,
a total of 156 electrons) are explicitly treated in the basis sets of 
triple-$zeta$ quality (double-$zeta$ for O atoms) \cite{Detalles}. 
In order to model the infinite lattice, a well-established approach has
been used, which consists in replacing the first-shell of neighbors
(in-plane and out-of-plane) by pseudopotentials, which incorporate
both electrostatic and exclusion effects of these ions, and in considering 
the Madelung field of the remote atoms of the periodic lattice, according to 
Evjen's technique \cite{Evjen}.

A restricted open-shell self-consistent field calculation (ROHF) for the 
quintet state has been carried out, which determines the four magnetic
orbitals ($a,b,c,d$ or their symmetry-adapted combinations corresponding to 
the irreducible representations $a_{1g}$, $b_{3u}$ and $e_u$ in the $D_{4h}$ 
symmetry group) (Figure 3). These four orbitals define a valence-space with 
one-electron and one-orbital per site, in one-to-one correspondence
with the model spaces of the Heisenberg Hamiltonian or the parent one-band 
Hubbard Hamiltonian.

The diagonalization of the valence CI matrix (CASCI), that is, a matrix
with dimension 36 in the delocalized basis set, gives a value of -28 meV
for the NN antiferromagnetic coupling, which is very far from the 
experimental estimation. This very limited CI only contains the Anderson
mechanism in the bare one-band model. This level of description misses
two important phenomena, namely intermediate charge-transfer from the oxygen 
atoms to the Cu atoms and the dynamical polarization effects of the internal
electrons and the surrounding atoms, which react to the fluctuation
of the field created by the active electrons. The treatment of these effects
requires much larger CI expansions.

In order to take  into account the first effect, namely the hopping between 
oxygen and Cu atoms, it is crucial to identify  the doubly-occupied orbitals
of the oxygen atoms which contribute to this mechanism. They are not necessarily
canonical orbitals, i.e. eigenstates of the Fock operator. The most-relevant 
ligand-centered orbitals will be obtained as energy-difference dedicated 
molecular orbitals \cite{Dedicados}.
These orbitals have been obtained as follow:\\
1.- From the four-electrons in four-orbital active space, a CI calculation
has been performed, limited to the single
excitations on the top of all the valence space determinants.\\
2.- The density matrices, $R_S$ and $R_Q$, for the lowest singlet and  
quintet states have been calculated. The excitation-energy
dedicated MOs are the eigenvectors of the difference of the density matrices
$R_S-R_Q$, restricted to the nearly doubly occupied MOs. The eigenvalues
of this matrix difference, called 'implication numbers', give  a measure of 
the participation of the corresponding orbital to the energy difference, 
hence to the antiferromangetic mechanism responsible for
the lowering ot the energy of the singlet state. The MOs  of largest
implication numbers are essentially spanned by $2p$ atomic orbitals of the
bridging oxygen atoms, as shown in Figure 4.\\
3.- Now these four orbitals will be added to the magnetic ones to define
an enlarged valence space involving 12 electrons in 8 MOs, corresponding to 
a two-band Hubbard  model since it includes both the $3d$-like orbital of 
the Cu atoms and $2p$ orbitals of the bridging oxygen directed along the
Cu-O bonds, with optimized delocalization tails on the external 
oxygen atoms. The effect of the dynamical polarization will be taken into 
account by performing all the single excitations on the top of this enlarged
valence space. The resulting CI vectors are expanded on a large space
($\sim$ 5 $\cdot$ 10$^6$ determinants). 

When applied to the dimeric cluster $Cu_2O_7$ the same strategy provides a 
value of $J$=128 meV, in good agreement with the experimental evaluations and
our previous CI estimates 138 meV\cite{Calzado:99a, Calzado:99b}, which
involved $d$ basis functions on the bridging oxygen atoms (which had to be
deleted here to make feasible the calculations on the plaquette).

The identification of the ab initio calculated spectrum of the tetrameric
clusters with the expected spacings of Figure 2 leads to the following
values of the effective interactions:\\
\begin{eqnarray}
h&=& 60. 22\:meV \:\:\: ; \:\: h'= 5.01 \: meV \nonumber\\
g_4&=& 7.00 \:meV\:\: \: ; \: \: g'_4= 0.49 \:meV \nonumber
\end{eqnarray}

From these values it is possible to establish the interactions as written
in the usual spin formulation of the four-body operator \cite{Eroles:99,
Lorenzana:99, Honda:93}:
\begin{eqnarray}
H&=&\sum_{<ij> NN} J (S_iS_j - \frac{1}{4}) + \sum_{<ij> NNN} 
J' ( S_iS_j - \frac{1}{4}) +  \nonumber \\
&+&K \sum_{<ijkl>}[ (S_iS_j)(S_kS_l)+(S_iS_l)(S_jS_k)-(S_iS_k)(S_jS_l)
- \frac{1}{16}]  \nonumber 
\end{eqnarray}
\\
where the higher multiplet energy is zero, $J$ corresponds to the NN 
interaction, $J'$ to the NNN coupling and $K$ to the four-spin cyclic exchange.
Notice that the last term produces the cyclic permutation of the four spins on the
plaquette plus ordinary two-spins exchanges of all the pairs of spins of the 
plaquette including those on the diagonals.  Written in the basis of the six 
S$_z$=0 determinants of the $abcd$ configuration, this Hamiltonian has the
following form:

\begin{tabular}{cccccc}
\cline{1-6}
$|a\bar{b}c\bar{d}|$ & $|\bar{a}b\bar{c}d|$ & $|ab\bar{c}\bar{d}|$ &
$|\bar{a}\bar{b}cd|$ & $|a\bar{b}\bar{c}d|$ & $|\bar{a}bc\bar{d}|$ \\
\cline{1-6}
$-2J$ &  $K/2$ & $J/2-K/8$ & $J/2-K/8$ & $J/2-K/8$  & $J/2-K/8$ \\
    & $ -2J$ & $J/2-K/8$ & $J/2-K/8$ & $J/2-K/8$  & $J/2-K/8$ \\
    &      & $ -J-J'$  & $ 0$      & $J'/2+K/8$ & $J'/2+K/8$ \\
    &      &         & $ -J-J'$  & $J'/2+K/8$ & $J'/2+K/8$ \\
    &      &         &         & $ -J-J'$   &  $ 0$     \\
    &      &         &         &          &  $-J-J'$   \\
\cline{1-6}
\end{tabular}
\\

Identifying the two matrices and omitting the negligible $g'_{4}$ term, 
one obtains:
\begin{eqnarray}
K&=& 2 g_{4}  \:\: \: \:\:\:\:\:\:\:\:\:\rightarrow\: K=14\: meV \nonumber\\
J&=& 2 h + \frac{K}{4} \:\: \:\rightarrow \:  J=124\: meV \nonumber \\
J'&=&2 h' -\frac{K}{4} \:\: \:\rightarrow \: J'=6.5\: meV \nonumber 
\end{eqnarray}

The value of the NN antiferromagnetic coupling ($J$) is in good agreement 
with both the previous estimation on the dimer ($J$=128 meV) and also with 
the experimental evaluations (128$\pm$6 meV\cite{Sulewski,Singh} 
and 134$\pm$5 meV\cite{Aeppli, Endoh, Hayden}).  Our estimate of the NNN
magnetic coupling ($J'$=6.5 meV) is in accord with the limit of 
$|J'|\le$9 meV, proposed for this compound from Raman experiments\cite{Hayden}. 
Concerning the four-spin cyclic exchange, experimental evaluations are not
available   and it is only possible to compare with the $K/J$ ratios used
in some recent numerical simulations of the absorption spectrum.
The here-presented values of $K$ and $J$ give $K/J \sim$ 0.11, which is lower
than the value of 0.25 assumed by Honda {\it et al.} {\cite{Honda:93} 
and than the value of 0.30 taken by Lorenzana {\it et al.}\cite{Lorenzana:99}
(from an earlier suggestion by Schmidt and Kuramoto \cite{Schmidt}) but larger
than the critical value, (K/J)$_c$=0.05 $\pm$0.04,
estimated by Sakai and Hasegawa \cite{Sakai:99} for the appearance of 
a magnetization plateau at half the saturation value in the $S=\frac{1}{2}$
antiferromagnetic spin ladders. The ratio of the NNN and NN interactions
is $J'/J$=0.051, somewhat  larger  than the value accepted by Lorenzana 
{\it et al.} ($J/J'$=0.04) \cite{Lorenzana:99}.

\section{Spin-phonon interactions}
Additional calculations have been performed to evaluate the dependence
of the  magnetic coupling constant on local geometrical
distortions of the lattice. This evaluation proceeds through
ab initio calculations on bimetallic clusters using the same strategy as
in the preceding section (same basis set, same kind of optimization of the
molecular orbitals and same type of Configuration Interaction calculations).

Five different local distortions have been considered, as shown in Figure 5.
Table 1 gives their corresponding force constants, associated frequency and
the derivative ${\Delta J}/{\Delta d_{Cu-O}}$. Concerning strongly localized
movements, these frequencies are different from the real frequencies of the
lattice, but offer an insight on the softness of the different motions.
Among the distortions, the movement of the bridging oxygen
atom along the Cu-Cu bond, lengthening one Cu-O bond and shortening the 
other one ($mode$ 2), has a small force constant ($\omega\sim$ 750 cm$^{-1}$),
but does not affect significantly the $J$ value.
The movements out of the Cu-Cu axis, either in-plane or along the $c$ axis
($modes$ 3, 4 and 5), induce strong changes on $J$ value  but the force 
constant and frequencies are large and, then these distortions do not seem
to be responsible for the dispersion of $J$. The movement
shortening (or lengthening) the Cu-Cu bonds ($mode$ 1) has both a significant 
impact on $J$ (${\Delta J}/{\Delta d_{Cu-O}}$=1700 cm$^{-1}\cdot$ \AA$^{-1}$)
and a low frequency ($\omega\sim$ 800 cm$^{-1}$).
 These values should be compared with those assumed in a recent
 work\cite{Eroles:99}, which takes  ${\Delta J}/{\Delta d_{Cu-O}}$=4350
 cm$^{-1}\cdot$ \AA$^{-1}$ and invokes the experimental frequencies 550 
 and 690 cm$^{-1}$. Our roughly calculated vibrational
 frequency is on line with the experimental one but the calculated dependence
 of $J$ on the Cu-O distance is half smaller that the value previously 
 proposed \cite{Eroles:99}.

\section{Conclusions}
This work has evaluated the amplitudes of the different interactions generally
invoked to explain the spectral features of $CuO_2$ layers which do not fit
with the simple Heisenberg Hamiltonian restricted to nearest neighbor 
coupling. Neither the spin-phonon coupling nor the next-nearest neighbor 
magnetic interactions nor the four-body cyclic effects are negligible, they
appear to be of the order of magnitude sometimes assumed in numerical 
simulations of the collective effects.  The here-presented ab initio
calculations are free from the simplifications of a one-band or even
of a two-band model Hamiltonian.
We believe that the so-obtained values of the generalized
distance-dependent Heisenberg Hamiltonian are reliable enough to deserve
to be used in the evaluation of the collective properties of the material.

\section*{Acknowledgements}
The authors are indebted to the European Commission for the TMR network
contract ERBFMRX-CT96-0079, Quantum Chemistry of Excited States.  
C.J.C. acknowledges the financial support through the TMR activity 
"Marie Curie research training grants" Grant No. HPMF-CT-1999-00285
established by the European Community.

\newpage
% Bibliography
\bibliographystyle{aip}

\newpage
\section*{Table 1}
Force constants ($K$), frequencies of vibration ($\omega$) and variation
of $J$ with the Cu-O distance (${\Delta J}/{\Delta d_{Cu-O}}$) of
different local distortions (see Figure 5)  in La$_2$CuO$_4$.
 
\begin{tabular}{cccc}
\cline{1-4}
mode & K(cm$^{-1}\cdot$\AA$^{-2}$) & $\omega$(cm$^{-1}$)
&${\Delta J}/{\Delta d_{Cu-O}}$ (cm$^{-1}\cdot$\AA$^{-1}$)\\
\cline{1-4}
$mode$ 1 &  6.53$\cdot$10$^5$  &  833 & -1693 \\
$mode$ 2 &  2.78$\cdot$10$^5$  &  763 &  $\sim$ 0  \\
$mode$ 3 &  8.69$\cdot$10$^5$(*) & 2549 & -1246 \\
$mode$ 4 &  6.76$\cdot$10$^5$(*) & 2246 & -1213 \\
$mode$ 5 &  2.45$\cdot$10$^6$(*) & 8568 & -2098 \\
\cline{1-4}
(*) K in cm$^{-1}\cdot rad^{-2}$
\end{tabular}
\newpage
\section*{Figure captions}
Figure 1. Fragment of the La$_2$CuO$_4$ lattice, containing the cluster
Cu$_4$O$_{12}$, which atoms are explicitly treated in the ab initio
calculations, and its first-shell of neighbor atoms, where pseudopotentials
have been placed to avoid an artificial polarization of the electronic
density of the terminal oxygen atoms.
\\
Figure 2. Spectrum of the plaquette, corresponding to an occupation of
one-electron per Cu site, written in the basis of the parameters of the
effective Hamiltonian. On the right, the symmetry of the different states in
the $D_{4h}$ group has been included.
\\
Figure 3. Linear combinations of the $3 dx^2-y^2$-type of orbitals,
containing non-negligible delocalization tails on the neighbor oxygen atoms.
These orbitals correspond to the $A_{1g}$ (a), $E_u$(b and c) and
$B_{3u}$ irreducible representation of symmetry in $D_{4h}$ group.
\\
Figure 4. Most-implicated dedicated molecular orbitals centered in the ligands.
These orbitals correspond to the $A_{1g}$ (a), $E_u$(b and c) and
$B_{3u}$ irreducible representation of symmetry in $D_{4h}$ group.
\\
Figure 5. Different local distortions in the Cu$_2$O$_9$ cluster. $Modes$ 1,
2 and 3 correspond to the distortions in the CuO$_2$ plane. $Mode$ 1 represents
the symmetric streching of the Cu-O$_{bridge}$ bond.
The Cu atoms have been symmetrically displaced along the $y$-axis.
$Modes$ 2 and 3 represent the displacement of the bridging oxygen atom
along the $y$ and the $x$ axis, respectively. $Mode$ 4 corresponds
to the movement of the central oxygen atom out of the $xy$ plane.
$Mode$ 5 represents a collective distortion, where the four oxygen atoms,
coordinated to one of the Cu atoms, go out of the plane.                         

\newpage
\begin{figure}[htb]
Figure 1. Calzado and Malrieu
\vbox to 5.0cm{
\includegraphics{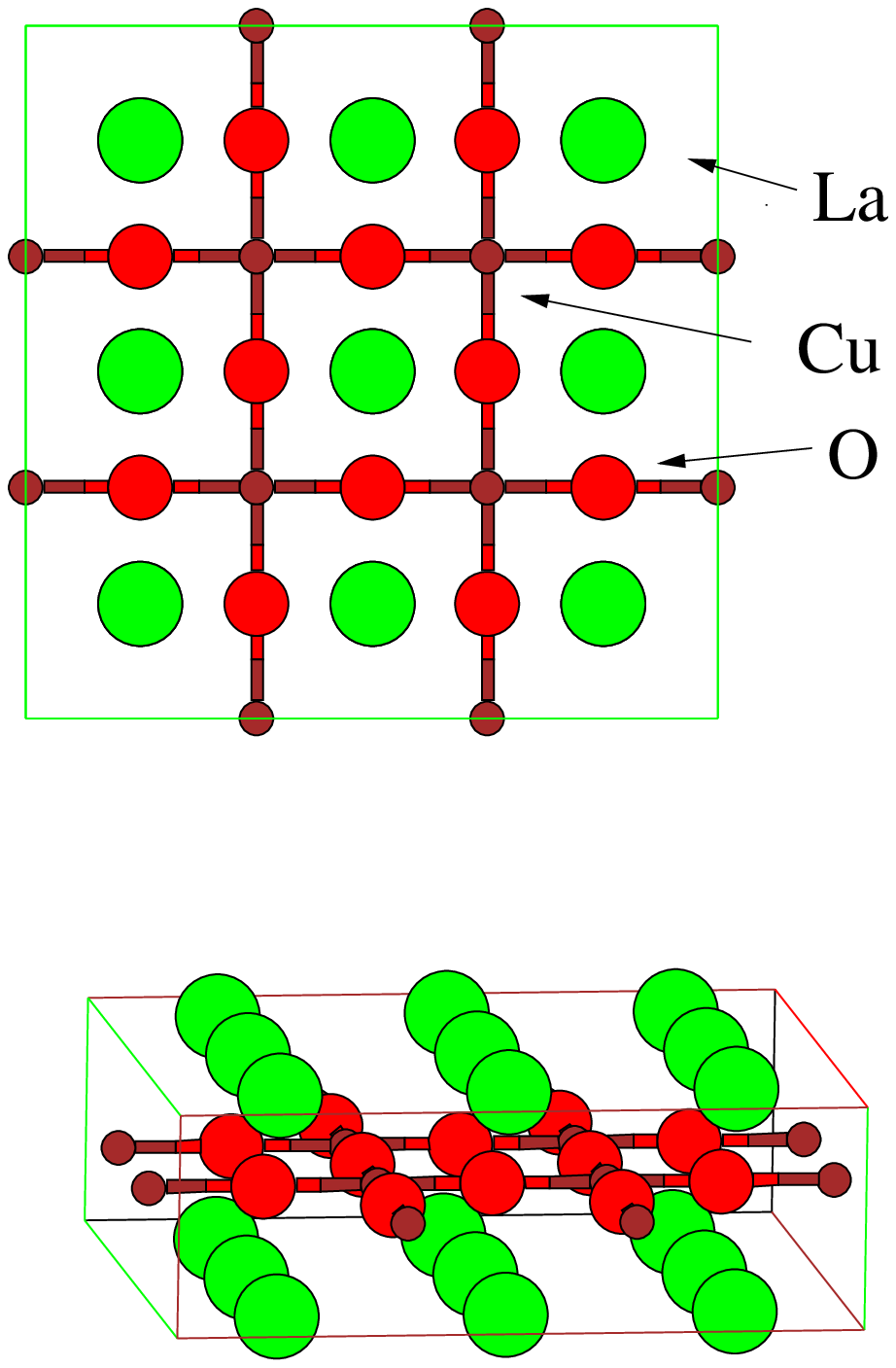}
\vfill}
\end{figure}

\newpage
\begin{figure}[htb] 
\vbox to 15.0cm{
\includegraphics{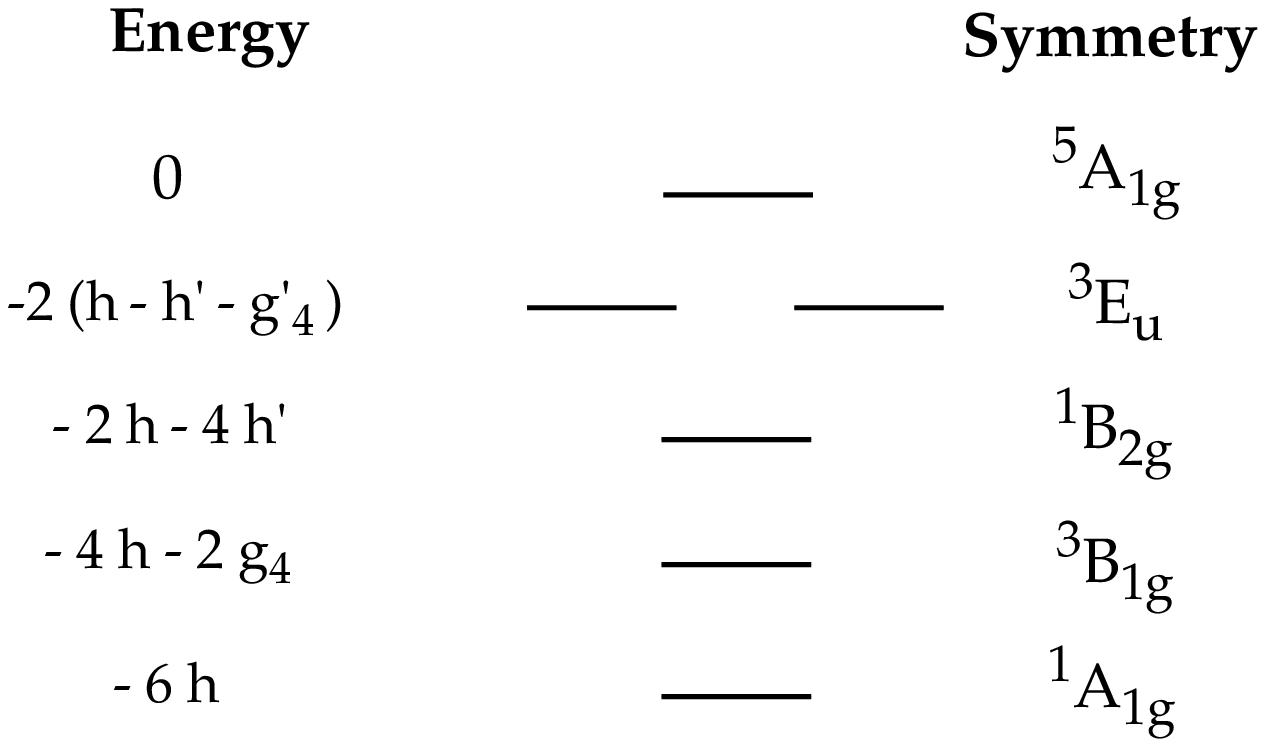}
\vfill}
\end{figure}
Figure 2. Calzado and Malrieu

\newpage
\begin{figure}[htb]
\vbox to 16.0cm{
\includegraphics{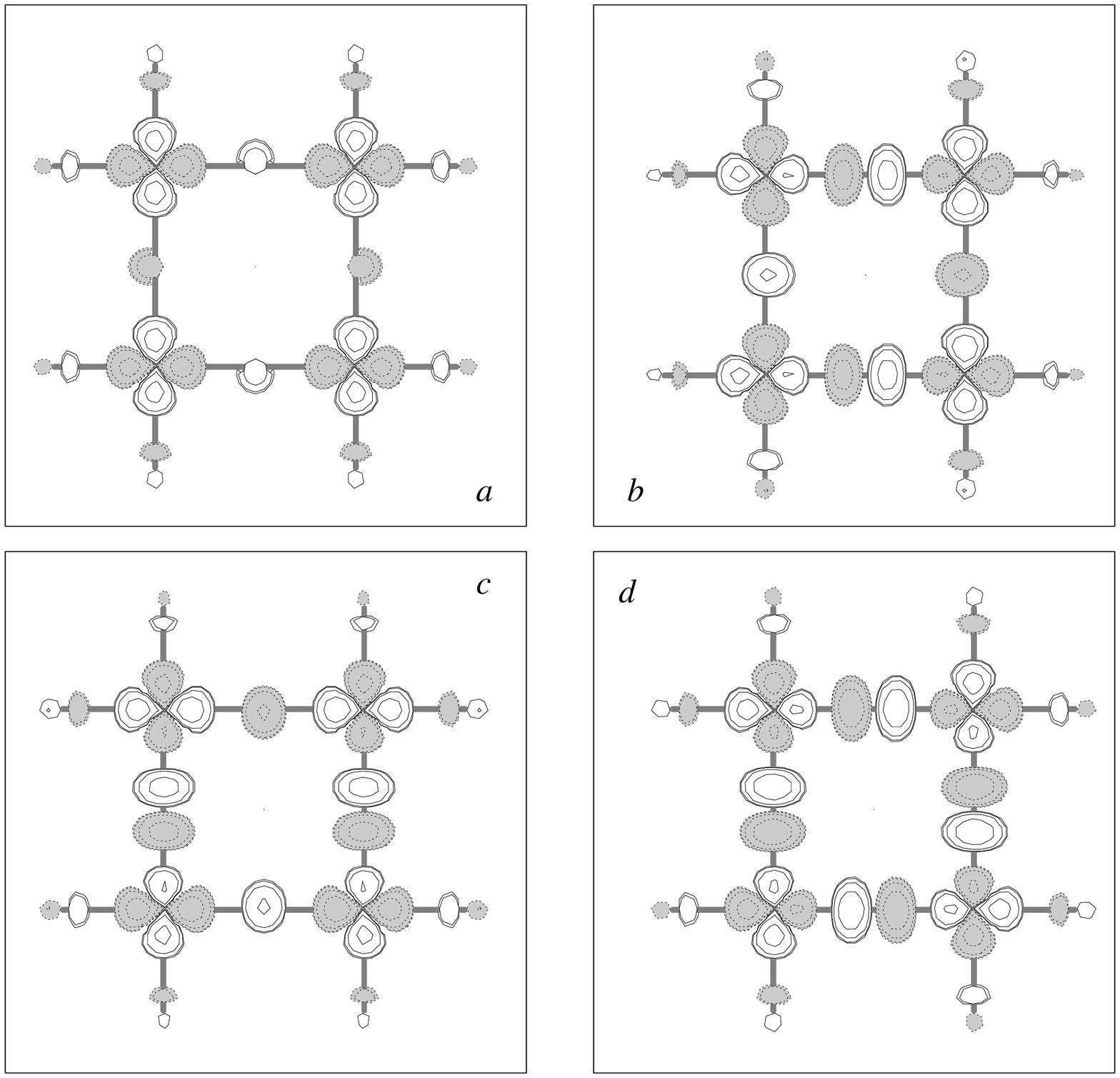}
\vfill}
\end{figure}
Figure 3. Calzado and Malrieu

\newpage
\begin{figure}[htb]
\vbox to 16.0cm{
\includegraphics{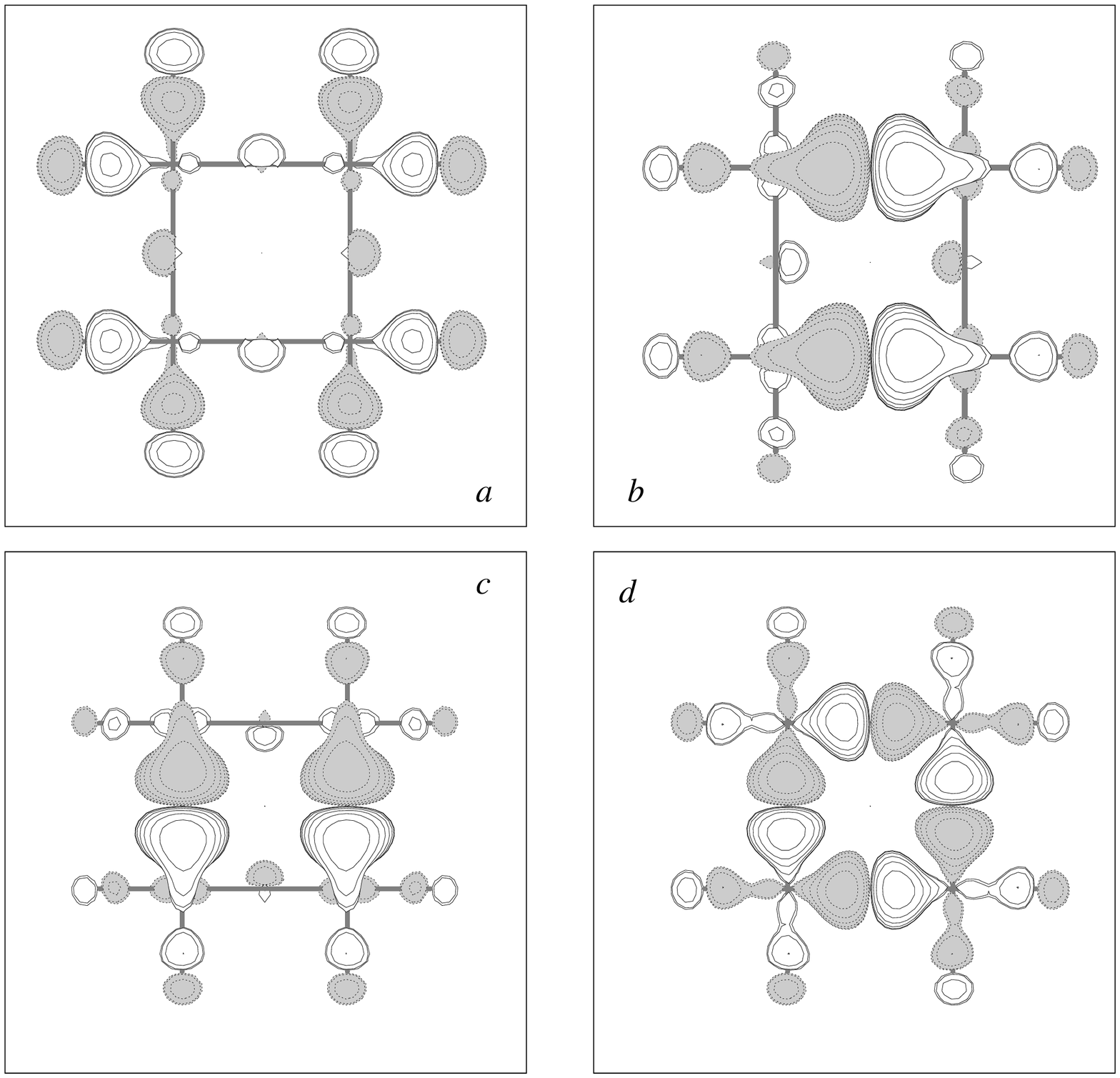}
\vfill}
\end{figure}
Figure 4. Calzado and Malrieu

\newpage
\begin{figure}[htb]
\vbox to 14.0cm{
\includegraphics{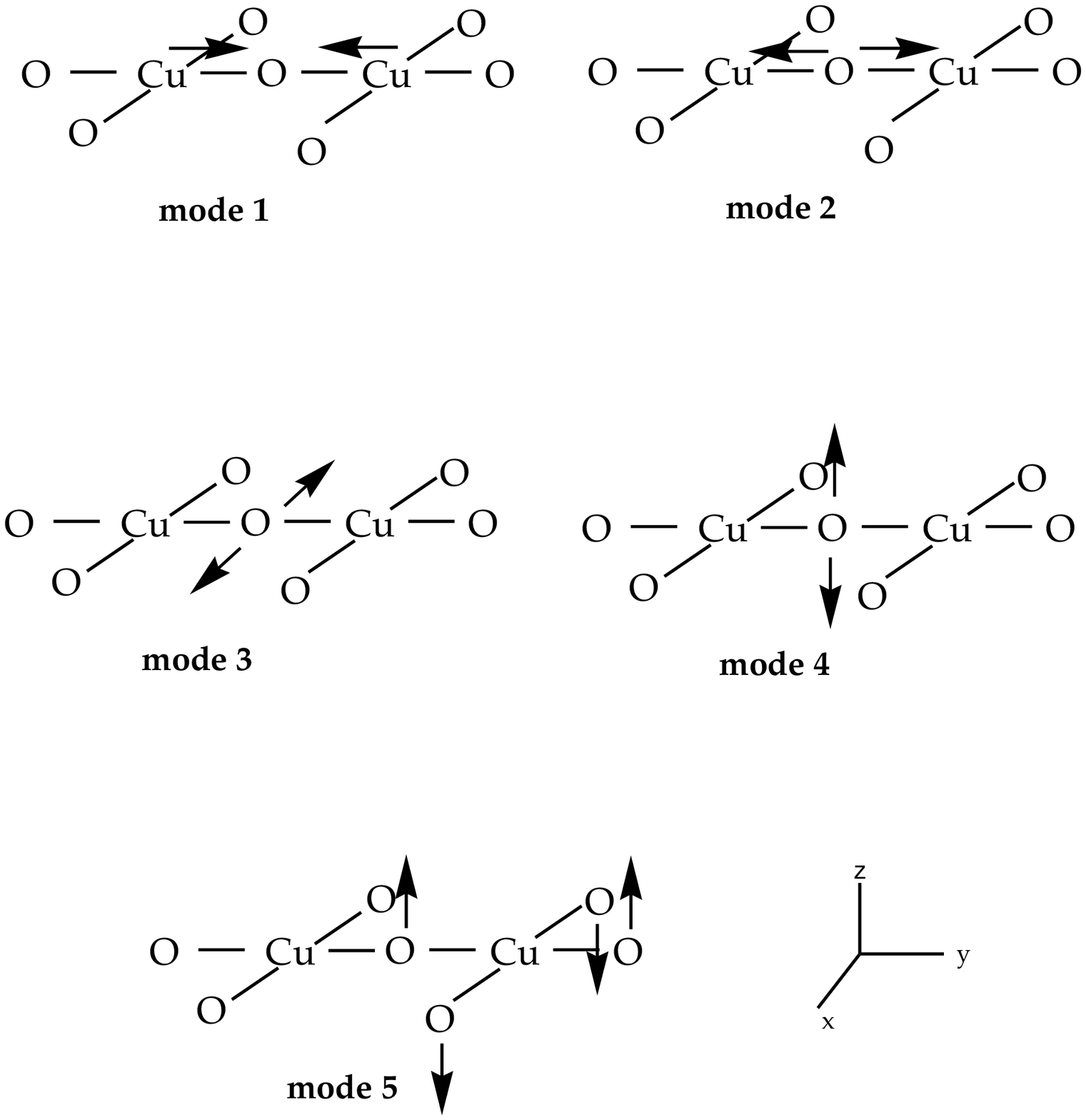}
\vfill}
\end{figure}
Figure 5. Calzado and Malrieu
\end{document}